# Minimum Distance and Convergence Analysis of Hamming-Accumulate-Acccumulate Codes

Alexandre Graell i Amat and Raphaël Le Bidan

*Abstract*—In this letter we consider the ensemble of codes formed by the serial concatenation of a Hamming code and two accumulate codes. We show that this ensemble is asymptotically good, in the sense that most codes in the ensemble have minimum distance growing linearly with the block length. Thus, the resulting codes achieve high minimum distances with high probability, about half or more of the minimum distance of a typical random linear code of the same rate and length in our examples. The proposed codes also show reasonably good iterative convergence thresholds, which makes them attractive for applications requiring high code rates and low error rates, such as optical communications and magnetic recording.

## I. INTRODUCTION

Applications such as magnetic recording or fiber-optic communications require error-correcting codes with a very high code rate ($R > 0.8$) and simple decoding algorithms amenable to high-throughput decoding architectures. Following the invention of turbo codes [1] and the rediscovery of Gallager's low-density parity-check (LDPC) codes [2], several high-rate low-complexity capacity-approaching codes have been proposed in the literature. Examples of such codes include turbo product codes (TPC) [3], the serial concatenation of a Hamming code with an accumulate code [4] or structured LDPC codes (see *e.g.* [5, chap. 17]). While such codes usually offer very good performance in the waterfall region, they are not asymptotically good in the sense that, unlike random codes, their minimum distance $d_{\min}$ does not grow linearly with block length, and thus may not be large enough to achieve the required error rates (*e.g.* of the order of $10^{-15}$ or lower for optical communications).

Recently, it was shown that repeat multiple-accumulate (RMA) codes with two or more accumulate stages are asymptotically good [6, 7]. It was further shown in [6] that high-rate code ensembles obtained by puncturing a low-rate repeat-accumulate-accumulate (RAA) code yield linear distance growth close to the Gilbert-Varshamov Bound (GVB). Unfortunately, as we shall see in Section IV, iterative decoding of punctured RMA does not converge for very high rates, making them impractical. In order to overcome this limitation, we consider in this work a class of high-rate double serially concatenated codes based on an outer, possibly extended Hamming code, with two accumulate codes. We first study

A. Graell i Amat is with the Department of Electronics, Institut TELECOM-TELECOM Bretagne, CS 83818 - 29238 Brest Cedex 3, France (e-mail: alexandre.graell@telecom-bretagne.eu). R. Le Bidan is with the Signal and Communications Department, Institut TELECOM-TELECOM Bretagne, CS 83818 - 29238 Brest Cedex 3, France (e-mail: raphael.lebidan@telecom-bretagne.eu). A. Graell i Amat is supported by a Marie Curie Intra-European Fellowship within the 6th European Community Framework Programme.

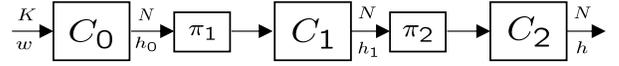

Fig. 1. Serial concatenation of an $(n, k)$ block code $C_0$ and two accumulators $C_1$ and $C_2$.

the ensemble-average finite-length weight enumerator for this code ensemble. Then, by generalizing the analytical tools introduced in [6, 7] to handle the case of an arbitrary outer linear block code, we study the asymptotic growth rate of the weight enumerator. We show, through selected examples, that the typical minimum distance of Hamming-accumulate-accumulate (HAA) codes grows linearly with block length, and provide a numerical estimate of the growth rate. Finally, we use extrinsic information transfer (EXIT) charts to estimate the iterative convergence thresholds. It is shown that the proposed codes have reasonably good convergence thresholds despite the double serial concatenation. Compared to TPCs much larger minimum distances can be achieved at the expense of a moderate loss in code convergence. Thus, HAA codes are a valid alternative when very high code rates and very low error rates are sought for.

## II. ENCODER STRUCTURE AND FINITE-LENGTH ENSEMBLE ENUMERATORS

We consider the code ensemble $\mathcal{C}$ formed by the serial concatenation of an $(n, k)$ outer block code $C_{\text{BC}}$ and two rate-1, memory-one, accumulate codes $C_1$ and $C_2$ with generator polynomials $g(D) = 1/(1 + D)$, connected through two interleavers $\pi_1$ and $\pi_2$. We assume that the interleaver size $N$ is a multiple $L$ of the length $n$ of the outer block code. The overall code rate is $R = K/N$, where $K = kL$ is the input block length and $N = nL$ is the output block length. We denote by $C_0$ the $(N, K)$ outer block code obtained by concatenating together $L$ successive codewords of $C_{\text{BC}}$. Trellis termination is used to transform the accumulate codes $C_1$ and $C_2$ into two equivalent $(N_l, K_l)$ block codes, $l = 1, 2$. The corresponding encoder is depicted in Fig. 1.

In this letter, we consider Hamming and extended Hamming (eHamming) codes for $C_{\text{BC}}$ since they can achieve minimum distances 3 and 4 with the highest code rate for a given dimension, with reasonable decoding complexity. A uniform distribution is assumed on the choice of the two permutations $\pi_1$ and $\pi_2$. In the following, the code ensemble $\mathcal{C}$ with an outer code of parameters $(n, k)$ will be referred to as the $(n, k)$AA ensemble.

Let $A_{w,h}^C$ denote the input-output weight enumerator (IOWE) of a block code $C$, *i.e.* the number of codewords with input weight $w$ and output weight $h$ in $C$. Similarly, let

$A_h^C = \sum_w A_{w,h}^C$ denote the weight enumerator (WE) of the code, *i.e.* the number of codewords of output weight $h$. Then, using the uniform interleaver concept [8,9], the ensemble-average WE of the code ensemble $\mathcal{C}$ can be computed as:

$$\overline{A}_h^{\mathcal{C}} = \sum_{h_0=0}^{N} \sum_{h_1=0}^{N} \frac{A_{h_0}^{C_0} A_{h_0,h_1}^{C_1} A_{h_1,h}^{C_2}}{\binom{N}{h_0}\binom{N}{h_1}} \quad (1)$$

The IOWE of an accumulate code with block length $N$ can be written in closed form as [10]:

$$A_{w,h}^{\text{ACC}} = \binom{N-h}{\lfloor w/2 \rfloor}\binom{h-1}{\lceil w/2 \rceil - 1} \quad (2)$$

In polynomial form, the WE of $C_0$ and $C_{\text{BC}}$ are linked by the following relationship [8]:

$$A^{C_0}(H) = \sum_{h=0}^{N} A_h^{C_0} H^h = [A^{C_{\text{BC}}}(H)]^L \quad (3)$$

Here, we derive another alternative expression of the WE of $C_0$ which is more convenient for asymptotic analysis. Consider a codeword $\mathbf{c}$ of output weight $h_0$ in $C_0$ and denote by $m_i$ the number of codewords of weight $i$ in $C_{\text{BC}}$ that participate in $\mathbf{c}$. Then, the WE of the outer block code, $A_{h_0}^{C_0}$, can be expressed as

$$A_{h_0}^{C_0} = \sum_{m_0,m_1,\ldots,m_n} \binom{L}{m_0, m_1, \ldots, m_n} \times (A_0^{C_{\text{BC}}})^{m_0} \cdots (A_n^{C_{\text{BC}}})^{m_n} \quad (4)$$

under the constraints $\sum_{i=0}^{n} i m_i = h_0$ and $\sum_{i=0}^{n} m_i = L$, and where $\binom{L}{m_0,m_1,\ldots,m_n} = \frac{L!}{m_0! m_1! \ldots m_n!}$ is the multinomial coefficient.

The ensemble-average WE can be used to bound the minimum distance $d_{\min}$ of the code ensemble $\mathcal{C}$. In particular, the following Proposition holds [11]:

*Proposition 1:* The probability that a code randomly chosen from an ensemble of linear codes $\mathcal{C}$ with average WE $\overline{A}_h^{\mathcal{C}}$ has $d_{\min} < d$ is upper bounded by

$$\Pr(d_{\min} < d) \leq \sum_{h=1}^{d-1} \overline{A}_h^{\mathcal{C}} \quad (5)$$

In Fig. 2 we display this probabilistic bound for four $(n,k)$AA code ensembles by plotting the largest weight $d$ in the right-hand side (RHS) of (5) yielding $\sum_{h=1}^{d-1} \overline{A}_h^{\mathcal{C}} < 1/2$, as a function of the code length $N$. Hence we expect at least half of the codes in $\mathcal{C}$ to have a minimum distance $d_{\min}$ at least equal to the value predicted by the curves. The considered outer codes are the $(31,26)$ and $(63,57)$ Hamming codes, as well as the $(32,26)$ and $(64,57)$ eHamming codes. A typical random linear code of length $N$ and rate $R$ has minimum distance $N\delta_{\text{GV}}$ [12], where $\delta_{\text{GV}}$ is the normalized Gilbert-Varshamov distance defined as the root $\delta \leq 1/2$ of the equation $\mathbb{H}_2(\delta) = 1-R$, and $\mathbb{H}_2(x)$ is the binary entropy function (with binary logarithm). For comparison purposes, we have also plotted in Fig. 2 the minimum distance predicted by the GVB for rates $R = 26/32$, $R = 26/31$ and $R = 57/63$ (the GVB for $R = 57/64$ is omitted for clarity). All codes appear to have a minimum distance that grows linearly with the block length. Furthermore, the achievable minimum distances are very high. For instance, the bound for the $(31,26)$AA ensemble predicts a typical minimum distance $d_{\min} \approx 117$ for block length $N = 8184$. For comparison, the product code $(128,120) \times (64,57)$ of similar code length and rate has $d_{\min} = 16$ only.

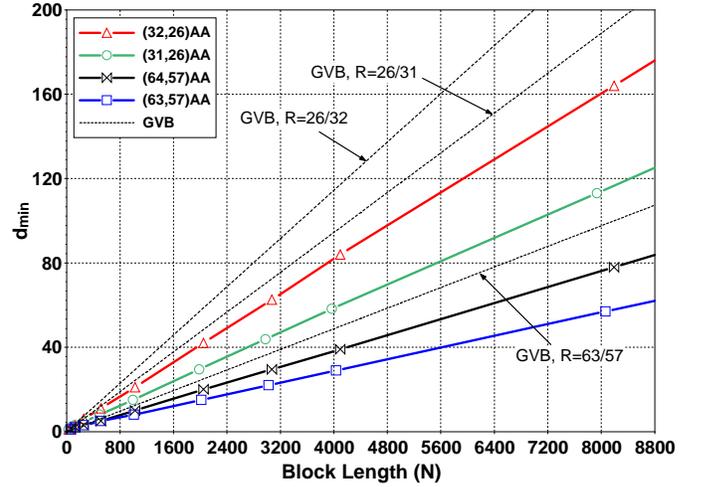

Fig. 2. Probabilistic bound on the minimum distance $d_{\min}$ versus block length $N$ for several $(n,k)$AA code ensembles.

## III. ASYMPTOTIC ENSEMBLE WEIGHT ENUMERATOR ANALYSIS

In this Section we analyze the asymptotic behavior of the WE to show that the minimum distance of the considered $(n,k)$AA code grows linearly with code length. To this end, we study the behavior of the *spectral shape* function of the code ensemble $\mathcal{C}$, defined as [2]

$$r(\delta) = \lim_{N \to \infty} \sup \frac{1}{N} \ln \overline{A}_{\lfloor \delta N \rfloor}^{\mathcal{C}} \quad (6)$$

where $\delta = h/N = h/(nL)$ is the normalized output weight. From (6) the WE can be expressed as $\overline{A}_h^{\mathcal{C}} \sim e^{Nr(\delta)}$. Therefore, if there exists some abscissa $\delta_{\min} > 0$ such that $\sup_{x \leq \delta} r(x) < 0 \quad \forall \delta < \delta_{\min}$, and $r(\delta) > 0$ for some $\delta > \delta_{\min}$, then it can be shown (using Proposition 1 for example) that, with high probability, the minimum distance of most codes in the ensemble grows linearly with the block length $N$, with growth rate $\delta_{\min}$ [7,13]. On the other hand, if $r(\delta)$ is equal to zero rather than strictly negative in the interval $(0, \delta_{\min})$, it cannot be concluded directly whether the minimum distance grows linearly or not with $N$ since the RHS in (5) may be bounded away from zero. As shown recently in [7], the spectral shape of RMA code ensembles exhibits such a behavior. However, based on WE analysis and appropriate bounding techniques, the authors of [7] were able to prove that for such code ensembles, the typical minimum distance indeed grows linearly with $N$, with growth rate $\delta_{\min}$.

### A. Spectral shape of the proposed code ensemble

Consider the $(N_l, K_l)$ block code $C_l$, $l = 0, 1, 2$. We define the asymptotic (logarithmic) behavior of the WE for $C_l$ as the function

$$a^{C_l}(\beta_l) = \lim_{N_l \to \infty} \sup \frac{1}{N_l} \ln A_{\lfloor \beta_l N_l \rfloor}^{C_l} \quad (7)$$

where $\beta_l = h_l/N_l$ is the normalized output weight. Similarly, we define the asymptotic behavior of the IOWE for $C_l$ as the function

$$a^{C_l}(\alpha_l, \beta_l) = \lim_{N_l \to \infty} \sup \frac{1}{N_l} \ln A^{C_l}_{\lfloor \alpha_l K_l \rfloor, \lfloor \beta_l N_l \rfloor} \quad (8)$$

where $\alpha_l = w_l/K_l$ is the normalized input weight. Then, using (1) and (7-8) in (6) and recalling Stirling's approximation for binomial coefficients $\binom{n}{k} \stackrel{n \to \infty}{\longrightarrow} e^{n \mathbb{H}_e(k/n)}$ where $\mathbb{H}_e(\cdot)$ is the binary entropy function with natural logarithms, the spectral shape function of the code ensemble $\mathcal{C}$ can be written as

$$\begin{aligned}
r(\delta) &= \lim_{N \to \infty} \sup \frac{1}{N} \ln \sum_{h_0=0}^{N} \sum_{h_1=0}^{N} \exp\{N a^{C_0}(\beta_0) \\
&\quad + N a^{C_1}(\beta_0, \beta_1) + N a^{C_2}(\beta_1, \beta) \\
&\quad - N \mathbb{H}_e(\beta_0) - N \mathbb{H}_e(\beta_1)\} \\
&\simeq \max_{0 \leq \beta_0, \beta_1 \leq 1} \{a^{C_0}(\beta_0) + a^{C_1}(\beta_0, \beta_1) \\
&\quad + a^{C_2}(\beta_1, \beta) - \mathbb{H}_e(\beta_0) - \mathbb{H}_e(\beta_1)\}
\end{aligned} \quad (9)$$

where $\beta_0 = h_0/N$ and $\beta_1 = h_1/N$. The last line follows from the well-known max-log approximation $\ln(e^a + e^b) \simeq \max(a, b)$ (see *e.g.* [14]).

The asymptotic behavior of the IOWE of the accumulate code $C_l$, $l = 1, 2$, is easily obtained by invoking again Stirling's approximation in (2), yielding [10],

$$a^{C_l}(\alpha_l, \beta_l) = (1 - \beta_l) \mathbb{H}_e\left(\frac{\alpha_l}{2(1 - \beta_l)}\right) + \beta_l \mathbb{H}_e\left(\frac{\alpha_l}{2\beta_l}\right) \quad (10)$$

The next Proposition addresses the problem of computing the asymptotic weight enumerator $a^{C_0}(\beta_0)$ for the outer block code $C_0$.

*Proposition 2:* Let $C_0$ be the $(N, K)$ block code obtained by concatenating together $L$ successive codewords of an $(n, k)$ block code $C_{\text{BC}}$. Let $p_i$ be the relative proportion of codewords of $C_{\text{BC}}$ of weight $i$ in a codeword of $C_0$, *i.e.* $p_i = m_i/L$. Define $\mathbf{P} = (p_0, p_1, \ldots, p_n)$ and $\mathbf{H}(\mathbf{P}) = -\sum_{i=0}^{n} p_i \ln p_i$, with the convention $0 \ln 0 = 0$. Then, the asymptotic IOWE of $C_0$ is given by the solution of the following convex optimization problem

$$a^{C_0}(\beta_0) = \max_{\mathbf{P}} \frac{1}{n}\left(\mathbf{H}(\mathbf{P}) + \sum_{i=0}^{n} p_i \ln A_i^{C_{\text{BC}}}\right) \quad (11)$$

under the constraints $\sum_{i=0}^{n} i p_i = n\beta_0$ and $\sum_{i=0}^{n} p_i = 1$.

*Proof:* The proof follows the approach proposed in [14] to obtain the spectral shape of generalized LDPC codes employing small Hamming codes at the check nodes. Since our problem is simpler, however, we arrive at a more tractable optimization problem (only the knowledge of the WE of $C_{\text{BC}}$ is required) and we avoid the conjectures made in [14].

Recalling first that $N_0 = N = nL$, we can express $a^{C_0}(\beta_0)$ in (7) as a function of $L$,

$$a^{C_0}(\beta_0) = \lim_{L \to \infty} \sup \frac{1}{nL} \ln A_{h_0}^{C_0} \quad (12)$$

with $\beta_0 = h_0/N_0 = h_0/nL$. Now, define the type $\mathbf{P} = (p_0, p_1, \ldots, p_n)$ of a codeword $\mathbf{c}$ in $C_0$, $\mathbf{c} = \mathbf{c}^1 \mathbf{c}^2 \ldots \mathbf{c}^L$, with $\mathbf{c}^j \in C_{\text{BC}}$, as the relative proportion of occurrences of codewords of $C_{\text{BC}}$ of weight $i$, $i = 0, 1, \ldots, n$, in $\mathbf{c}$ [15]. The set of all length-$L$ sequences each containing $m_i$ occurrences of codewords of $C_{\text{BC}}$ with weight $i$ is called the type class of $\mathbf{P}$, denoted $T(\mathbf{P})$. It follows that $|T(\mathbf{P})| = \binom{L}{m_0, m_1, \ldots, m_n}$. From [15, Thm. 11.1.3] we have that

$$|T(\mathbf{P})| \stackrel{L \to \infty}{\longrightarrow} e^{L \mathbf{H}(\mathbf{p})} \quad (13)$$

Finally, using (13) in (4) and (12) we obtain:

$$\begin{aligned}
&a^{C_0}(\beta_0) \\
&= \lim_{L \to \infty} \sup \frac{1}{nL} \ln \sum_{p_0, \ldots, p_n} e^{L \mathbf{H}(\mathbf{p})} (A_0^{C_{\text{BC}}})^{p_0 L} \cdots (A_n^{C_{\text{BC}}})^{p_n L}
\end{aligned} \quad (14)$$

from which (11) follows. ∎

The asymptotic WE (11) admits a closed-form expression for a few simple codes such as the $(nK, K)$ block repetition code of rate $R = 1/n$.

*Example 1: Asymptotic WE of the $(nK, K)$ repetition code*

Let $C_0$ be the $(nK, K)$ block code formed by concatenating together $K$ codewords of an $(n, 1)$ repetition code $C_{\text{BC}}$. We have $A_0^{C_{\text{BC}}} = A_n^{C_{\text{BC}}} = 1$ and $A_i^{C_{\text{BC}}} = 0$ for $i = 1, \ldots, n-1$. Define $p_0 = m_0/K$ and $p_n = m_n/K$. Using (11) we obtain:

$$\begin{aligned}
a^{C_0}(\beta_0) &= \max_{\mathbf{P}} \frac{1}{n}\left(\mathbf{H}(\mathbf{P}) + \sum_{i=0}^{n} p_i \ln A_i^{C_{\text{BC}}}\right) \\
&= \max_{p_0, p_n} \frac{1}{n}(-p_0 \ln p_0 - p_n \ln p_n)
\end{aligned} \quad (15)$$

under the two constraints $np_n = n\beta_0$ and $p_0 + p_n = 1$. Therefore,

$$a^{C_0}(\beta_0) = \frac{1}{n}(-(1-\beta_0)\ln(1-\beta_0) - \beta_0 \ln \beta_0) = \frac{1}{n} \mathbb{H}_e(\beta_0) \quad (16)$$

which is a well-known result (see *e.g.* [6, 10]).

For more general block codes with known WE, $a^{C_0}(\beta_0)$ can be evaluated numerically using standard convex optimization software.

### B. Asymptotic growth rate of the minimum distance

The numerical evaluation of (9) for the $(31, 26)$AA, $(32, 26)$AA, $(63, 57)$AA and $(64, 57)$AA codes ensembles is shown in Fig. 3. We have also plotted the spectral shape $r(\delta) = \mathbb{H}_e(\delta) - (1 - R) \ln(2)$ for the corresponding random linear code ensembles. The behavior of the spectral shape of the $(n, k)$AA ensembles is similar to the one obtained in [7] for RMA codes. It is strictly positive in the range $(\delta_{\min}, 1 - \delta_{\min})$ for some $0 < \delta_{\min} < 1/2$, and zero elsewhere. Since a closed form expression of the WE of the outer block code $C_0$ is not available, in contrast to RMA codes [7] we cannot provide a formal proof that the considered code ensembles are asymptotically good. However, extensive numerical experiments using (5) show that $\Pr(d_{\min} < \lfloor \delta_{\min} N \rfloor) \longrightarrow 0$ as $N$ gets large, which suggests that the results of [7] hold true for more general outer codes than just repetition codes. In Table I we report the estimated values of $\delta_{\min}$ for several $(n, k)$AA ensembles based on high-rate Hamming and extended Hamming codes. For comparison purposes, the normalized minimum distance $\delta_{\text{GV}}$ of random linear codes is also given. The estimated asymptotic growth rates $\delta_{\min}$ are in good agreement with the slope of the curves obtained by the finite-length analysis in Section II. Hence we conclude that HAA codes have minimum distance growing linearly with block length as $\delta_{\min} N$.

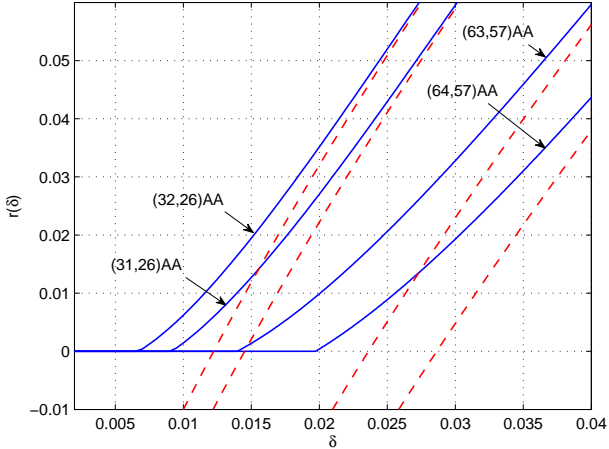

Fig. 3. Spectral shape of selected $(n,k)$AA code ensembles. The spectral shape of the corresponding random linear codes of same rate is also shown in dashed lines.

TABLE I
ASYMPTOTIC NORMALIZED MINIMUM DISTANCE $\delta_{\min}$ AND ITERATIVE CONVERGENCE THRESHOLD FOR SELECTED $(n,k)$AA CODE ENSEMBLES.

| Code | $\delta_{\min}$ | $\delta_{\text{GV}}$ | Threshold | Constrained capacity |
|---|---|---|---|---|
| $(32,26)$AA | 0.0197 | 0.0286 | 3.34 dB | 2.14 dB |
| $(31,26)$AA | 0.0140 | 0.0236 | 3.48 dB | 2.39 dB |
| $(64,57)$AA | 0.0091 | 0.0145 | 4.10 dB | 3.03 dB |
| $(63,57)$AA | 0.0067 | 0.0122 | 4.20 dB | 3.26 dB |
| $(128,120)$AA | 0.0042 | 0.0073 | 4.70 dB | 3.93 dB |
| $(127,120)$AA | 0.0032 | 0.0063 | 4.79 dB | 4.11 dB |

## IV. CONVERGENCE ANALYSIS

In this section, we investigate the iterative convergence behavior of $(n,k)$AA code ensembles by means of an EXIT chart analysis [16]. The iterative convergence thresholds for the considered $(n,k)$AA codes ensembles are given in Table I. BPSK transmission over an AWGN channel is assumed. Convergence is achieved within 0.7 dB–1.2 dB from the corresponding Shannon limit. Furthermore, the thresholds get closer to capacity for higher rates. For instance, the predicted thresholds for the $(31,26)$AA and the $(63,57)$AA code ensembles are 3.48 dB and 4.20 dB, respectively, *i.e.* 1.09 dB and 0.94 dB away from the constrained capacity. We also note that Hamming outer codes may be preferable to extended Hamming outer codes in practice since the former provide better thresholds at higher code rate and with smaller decoding complexity (number of states reduced by half in the code trellis), at the expense of a slightly smaller asymptotic growth rate of the minimum distance.

Comparison of the proposed codes with Hamming-accumulate codes, denoted hereafter as $(n,k)$A codes, show that the latter have better convergence thresholds. For example, the $(31,26)$A and $(63,57)$A codes converge at 2.81 dB and 3.58 dB, respectively. Thus double concatenation incurs a loss of 0.6 dB with respect to single concatenation. On the other hand, Hamming-accumulate codes (and more generally BCH-accumulate codes [17, 18]) are asymptotically bad.

TABLE II
NORMALIZED MINIMUM DISTANCE $\delta_{\min}$ FOR SELECTED RANDOMLY-PUNCTURED $R_3$AA CODES WITH TARGET RATE $R$.

| $R$ | $\delta_{\min}$ | $\delta_{\text{GV}}$ |
|---|---|---|
| 26/32 | 0.0282 | 0.0286 |
| 26/31 | 0.0233 | 0.0236 |
| 57/64 | 0.0143 | 0.0145 |
| 57/63 | 0.0120 | 0.0122 |
| 120/128 | 0.0072 | 0.0073 |
| 120/127 | 0.0062 | 0.0063 |

It was shown in [6] that the ensemble of randomly-punctured RAA codes is asymptotically good and achieves linear minimum distance growth close to the GVB, even for very high rates. As an example, we have reported in Table II the normalized minimum distance $\delta_{\min}$ for punctured RAA codes with repetition factor 3 (hereafter denoted as $R_3$AA) and same rate than the $(n,k)$AA codes considered in Table I. The results show that punctured $R_3$AA codes significantly outperform HAA codes in terms of normalized minimum distance. On the other hand, as noted in [6, 11], the convergence thresholds of RAA under iterative decoding are generally away from capacity. The situation gets even worse in the presence of puncturing. Our experiments have shown the existence of a maximum rate $R_{\max}$ above which the EXIT chart for randomly-punctured RAA codes shows an early cross between the two EXIT curves, even at very high (ultimately infinite) $E_b/N_0$, meaning that iterative decoding cannot converge. This limiting rate is around $R_{\max} \approx 0.695$ for punctured $R_3$AA codes. Therefore iterative decoding of randomly-punctured $R_3$AA codes do not converge at the code rates considered in Tables I and II. Simulations confirmed this prediction. Note that lowering the repetition factor does not help. For example, we found that $R_{\max} \approx 0.42$ for $R_6$AA.

In [19] a family of asymptotically good rate-compatible protograph-based LDPC codes was introduced which supports any code rate of the form $R = (n+1)/(n+2)$ for $n = 0, 1, 2, \ldots$ These codes, nicknamed $AR_4$JA, combine rate flexibility with excellent convergence thresholds, within 0.45 dB or less from the constrained capacity at all rates. For instance, $AR_4$JA codes converge at 0.37 dB from capacity at rate $R = 5/6$ [19]. This is 0.75 dB better than $(31, 26)$AA codes. On the other hand, the minimum distance growth rate for $AR_4$JA codes is only $\delta_{\min} = 0.015$ for the lowest code rate $R = 1/2$ [19], whereas HAA codes achieve almost the same growth rate ($\delta_{\min} = 0.014$) but at much higher code rate $R = 26/31 \approx 5/6$. Thus, HAA codes are expected to achieve significantly higher minimum distance than $AR_4$JA codes of same rate and code length.

## V. SIMULATION RESULTS

We compared the performance of the proposed $(n,k)$AA codes with that of TPCs, structured LDPC codes and $(n,k)$A codes, assuming BPSK modulation and transmission over an AWGN channel. In all simulations, random interleavers and a maximum of 30 decoding iterations were considered for $(n,k)$AA codes and $(n,k)$A codes. Turbo decoding of the $(n,k)$AA codes was realized as described in [9]. An optimum trellis-based MAP decoder was used to decode the Hamming codes. Turbo decoding of the TPC was realized using the Chase-Pyndiah decoding algorithm [3] with 16 test patterns and a maximum of 16 iterations (no significant improvement was observed beyond). One iteration comprises here a row-decoding step followed by a column-decoding step. Note that the Chase-Pyndiah decoder may also be applied to decode the outer block code in the $(n, k)$A and $(n, k)$AA concatenated schemes, resulting in lower complexity with performance very close to MAP decoding [18].

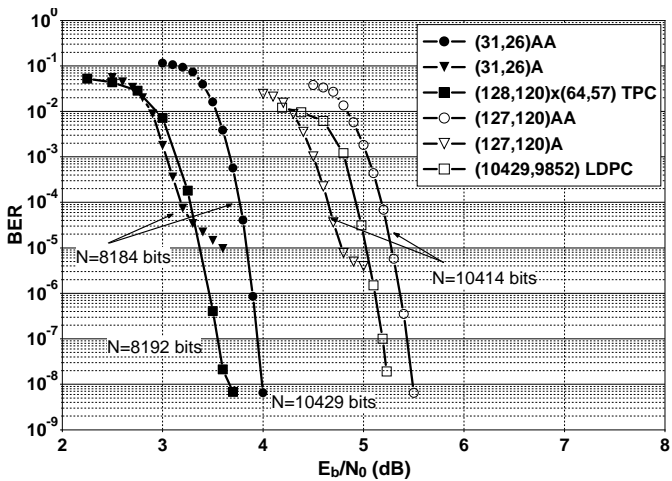

Fig. 4. Bit error probability performance for $(31, 26)$AA and $(127, 120)$AA ensembles on an AWGN channel with BPSK modulation.

In Fig. 4 the bit error rate (BER) performance of the $(31, 26)$AA code is compared with the performance of the $(128, 120) \times (64, 57)$ TPC and $(31, 26)$A code, respectively. The three codes have similar rate. The block length is $N = 8184$ for the $(31, 26)$AA and $(31, 26)$A codes and $N = 8192$ for the TPC. The simulated curve for the $(31, 26)$AA code is in agreement with the convergence threshold predicted by the EXIT charts. A loss of $\sim 0.6$ dB and of $\sim 0.5$ dB are observed with respect to the $(31, 26)$A code and the TPC, respectively. However the $(31, 26)$AA code is expected to yield significantly lower error floor thanks to a much higher $d_{\min}$. For instance, the minimum distance of the TPC is 16 while the majority of the codes in the $(31, 26)$AA ensemble are expected to have minimum distance $\sim 114$ with high probability.

The performance of the $(127, 120)$AA code is also given in Fig. 4. It is compared with the performance of the $(127, 120)$A code and also with a $(10429, 9852)$, rate-$R = 0.945$, graph-theoretic LDPC code described in [5, Fig 17.43]. The latter was selected because of its code rate $R \approx 120/127$ and its good performance under iterative decoding for a structured LDPC code (its Tanner graph is free of length-4 cycles). Accordingly, the block length was set to $N = 10414$ bits for both the $(127, 120)$AA and the $(127, 120)$A codes. The $(127, 120)$A code performs the best in the waterfall region. However, it shows the highest error floor due to a poor minimum distance (the floor would be lowered about 2 decades by using a carefully optimized interleaver). The $(127, 120)$AA code shows a loss of $\sim 0.5$ dB and $\sim 0.25$ dB in the waterfall region with respect to the $(127, 120)$A code and the structured LDPC code, respectively. The minimum distance of the structured LDPC code is not known precisely but is guaranteed to be at least 7 [5, p. 934]. On the other hand, at least half of the codes in the $(127, 120)$AA code ensemble have minimum distance 33. The latter are therefore expected to perform better at very low error rates.

## VI. CONCLUSIONS

We studied the serial concatenation of a Hamming code with two accumulate codes and showed that the resulting code ensemble is asymptotically good in the sense that most codes in the ensemble have minimum distance growing linearly with block size. We described a computational method to estimate the asymptotic growth rate of the minimum distance. Although only Hamming codes were considered here for the purpose of designing high-rate codes, the proposed method naturally extends to other outer linear block code. Finally, an EXIT chart analysis showed that Hamming-accumulate-accumulate codes exhibit reasonably good iterative convergence thresholds in spite of the double serial concatenation.

## ACKNOWLEDGMENTS

The authors would like to thank Chiara Ravazzi, from Politecnico di Torino, for helpful discussions. They also thank Karine Amis, from TELECOM Bretagne, for providing us the simulation results for the turbo product codes in Section V.